\documentclass[aps,pre,reprint,amsmath,amssymb,superscriptaddress,showpacs]{revtex4-1}

\usepackage{mathtools}
\usepackage{amsmath}    % need for subequations
\usepackage{graphicx}   % need for figures
\usepackage{verbatim}   % useful for program listings
\usepackage{color}      % use if color is used in text
\usepackage{lipsum}
\usepackage{subfigure}  % use for side-by-side figures
\usepackage{hyperref}   % use for hypertext links, including those to external documents and URLs

\begin{document}

%\preprint{}

%\title{Nonlinear dynamics and chaos in one dimensional disordered lattices}
\title{Characteristics of chaos evolution in one-dimensional disordered nonlinear lattices}

%\author{Name}
%\thanks{}

\author{B.~Senyange}
\affiliation{Department of Mathematics and Applied Mathematics, University of Cape Town, Rondebosch, 7701, Cape Town, South Africa}
\author{B.~Many Manda}
\affiliation{Department of Mathematics and Applied Mathematics, University of Cape Town, Rondebosch, 7701, Cape Town, South Africa}
\author{Ch.~Skokos}
\email{haris.skokos@uct.ac.za}
\thanks{Corresponding author.}
\affiliation{Department of Mathematics and Applied Mathematics, University of Cape Town, Rondebosch, 7701, Cape Town, South Africa}
\affiliation{Max Planck Institute for the Physics of Complex Systems, N\"othnitzer Str.~38, D-01187 Dresden, Germany}

\date{\today}

%\keywords{}

\begin{abstract}
We numerically investigate the characteristics of chaos evolution during wave packet spreading in two typical one-dimensional nonlinear disordered lattices: the Klein-Gordon system and the discrete nonlinear Schr\"{o}dinger equation model. Completing previous investigations \cite{SGF13} we verify that chaotic dynamics is slowing down both for the so-called `weak' and `strong chaos' dynamical regimes encountered in these systems,  without showing any signs of a crossover to regular dynamics. The value of the finite-time maximum Lyapunov exponent $\Lambda$ decays in time $t$ as $\Lambda \propto t^{\alpha_{\Lambda}}$, with $\alpha_{\Lambda}$ being different from the $\alpha_{\Lambda}=-1$ value observed in cases of regular motion. In particular, $\alpha_{\Lambda}\approx -0.25$ (weak chaos) and $\alpha_{\Lambda}\approx -0.3$ (strong chaos) for both models, indicating the dynamical differences of the two regimes and the generality of the underlying chaotic mechanisms. The spatiotemporal evolution of the deviation vector associated with $\Lambda$ reveals the meandering of chaotic seeds inside the wave packet, which is needed for obtaining the chaotization of the lattice's excited part.
\end{abstract}


\maketitle

%==============================
\section{Introduction}
\label{sec:introducion}

Disordered systems are spatially extended models of many degrees of freedom trying to mimic heterogeneity in nature. Typically they are obtained by attributing to one of the system's parameters a different, random value for each degree of freedom. Such systems offer a perfect test bed for understanding the dynamical properties of multidimensional Hamiltonian models, while at the same time they are of significant practical interest as they can be used for describing several important physical processes like for example the conductivity of materials, the propagation of light in optical waveguides, the dynamics of Bose-Einstein condensates, the structural behavior of granular solids and the dynamics of DNA molecules.

It is well-known that in linear disordered systems energy excitations remain localized. This phenomenon was first theoretically studied by Anderson in 1958 \cite{A58} (and for this reason it is called `Anderson localization'), and afterwards it was also observed experimentally \cite{WBLR97,CSG00,RZ03,GC05,SGAM06,BJZBHLCSBA08,HSPSV08,KMZD11}. The effect of nonlinearity in disordered systems has attracted extensive attention in the last decade, in theory and simulations \cite{KKFA08,PS08,FKS09,SKKF09,GS09,VKF09,MAPS09, MP10,LBKSF10,SF10,KF10,F10,JKA10,B11,BLSKF11,BLGKSF11,A11,ILF11, MLT12,VG12,MF12,B12,MP12,LBF12,MI12,VG13b,MP13,SGF13,LTDMIM13,LBF13, TSL14,ILF14,ABSD14,ES14,LIF14,B14,F15,MKP16,ATS16,ASB17,I17,ATS18},
as well as in experiments \cite{SBFS07,RDFFFZMMI08,LAPSMCS,LDTRZMLDIM11}. A fundamental question in this context is what happens to energy localization in the presence of nonlinearities.

Extensive numerical studies of the effect of nonlinearity on the propagation of initially localized energy excitations in disordered variants for two typical one-dimensional Hamiltonian lattice models, namely the Klein-Gordon (DKG) oscillator chain and the discrete nonlinear Schr\"{o}dinger (DDNLS) equation, determined the statistical characteristics of energy spreading and showed that nonlinearity destroys localization \cite{FKS09,SKKF09,LBKSF10,SF10,F10,BLSKF11,BLGKSF11,SGF13}. In those papers the existence of different dynamical spreading regimes, namely the so-called `weak' and `strong chaos'  regimes, was revealed, their particular dynamical characteristics were determined and their appearance was theoretically explained. In particular, it was theoretically predicted and numerically verified that nonlinearity leads to the subdiffusive spreading of wave packets in accordance to the observations of \cite{M98,KKFA08,PS08,GS09}. More specifically, it was shown that in the case of one-dimensional lattices the wave packet's second moment $m_2$ grows in time $t$ as
$m_2 \propto t^a$, with $a=1/3$ and $a=1/2$ for the weak and strong chaos regimes respectively.
A physical mechanism of this subdiffusion in the DDNLS model has
been suggested in \cite{MI12,I17} where the exponent $a=1/3$ has been explained as well. Experimental evidences of such subdiffusive spreadings in Bose-Einstein condensates were provided in \cite{LDTRZMLDIM11}. Subdiffusive spreading was  also numerically observed for two-dimensional disordered lattices \cite{GS09,LBF12,LIF14}.

Although, nowadays is common knowledge that energy spreading in disordered lattices is a chaotic process, the characteristics of this chaotic behavior have not been studied in detail. The first attempt to systematically investigate chaos in one-dimensional disordered, nonlinear lattices was performed in \cite{SGF13} where the chaotic wave packet spreading in the weak chaos spreading regime of the DKG model was studied in detail. For that particular case it was shown that although chaotic dynamics slows down, it does not cross over into regular dynamics. In addition, that work provided some first numerical evidences on how chaotic behavior appears in disordered lattices by indicating that `chaotic hot spots', where few lattice sites seem to behave more chaotically than others, meander through the system as time evolves sustaining its chaoticity.

In \cite{SGF13} the computation of the most commonly used chaos indicator, the finite-time maximum Lyapunov  exponent  $\Lambda$ \cite{BGGS80a,BGGS80b,S10}, was used to verify the DKG system's chaoticity in the weak chaos regime. It was found that, as the number of lattice's excited degrees of freedom increases when the energy spreads to more lattice sites, $\Lambda$ decreases in time $t$ following the power law $\Lambda \propto t^{-0.25}$, which is different from the behavior $\Lambda \propto t^{-1}$ observed in the case of regular motion. Thus, the system becomes less chaotic, while  the dynamics does not  show any tendency to crossover to regular behavior (at least up to the computationally  accessible times) as it was speculated in \cite{JKA10,A11}.

Chaoticity by itself is not enough to guarantee thermalization of disordered systems \cite{TSL14} and support subdiffusion theories. The needed, additional ingredient is the spatiotemporal fluctuations of the chaotic seeds inside the excited part of the lattice, something which was shown in  \cite{SGF13} through the time evolution of the deviation vector (i.e.~the displacement from the studied orbit in the system's phase space) used for the computation of $\Lambda$. Since this vector eventually aligns with the most unstable direction in the   system's phase space the time evolution of its coordinates showed that localized chaotic seeds meander through the wave packet contributing in this way to its thermalization.

In the present paper we extend these investigations by considering not only the weak chaos spreading regime but also strong chaos cases, in order to identify possible similarities or differences in the way chaos evolves in these regimes. By performing extensive numerical computations of $\Lambda$, as well as of the related deviation vector distributions (DVDs), we  investigate the characteristics of chaoticity in detail. We  perform our investigations not only for the DKG model (completing in this way the study of \cite{SGF13}) but also for the DDNLS in order to  verify the generality of our findings.

The paper is organized as follows: in Sect.~\ref{sec:models} we present the two Hamiltonian models we consider in our study and provide information about the numerical tools we use in our investigations: computed quantities, integration techniques etc. In Sect.~\ref{sec:num} we present our numerical findings about the chaotic behavior of the DKG and the DDNLS systems for various parameter cases emphasizing the computation of the finite time  maximum Lyapunov  exponent $\Lambda$ and the corresponding DVDs. Finally in Sect.~\ref{sec:sum} we summarize our results and discuss their significance.

%==============================
\section{Models and computational methods}
\label{sec:models}

In our study we consider two Hamiltonian models of one-dimensional nonlinear disordered lattices. The first one is the quartic DKG lattice chain of $N$ oscillators described by the Hamiltonian function
\begin{equation}
H_K = \sum_{l = 1}^{N} \frac{p_l ^2}{2} + \frac{\tilde{\epsilon}_l }{2} q_l ^2 +
\frac{q_l ^4}{4} + \frac{1}{2W}\left( q_{l+1} - q_{l} \right) ^2,
\label{eq:H_KG}
\end{equation}
where $q _l$ and $p_l$ respectively represent the generalized position and momentum of site $l$, $\tilde{\epsilon}_l$ are disorder parameters of the on-site potential whose values are  uniformly chosen from  the interval $\left[\frac{1}{2}, \frac{3}{2} \right]$ and $W$ is the
disorder strength. The corresponding equations of motion are
\begin{equation}
\ddot{q}_l = - \left[ \tilde{\epsilon}_l q_l + q_l ^3 + \frac{1}{W}\left(2q_l - q_{l-1} -
q_{l + 1} \right)  \right].
\label{eq: eq of motion for the KG general}
\end{equation}
The Hamiltonian function (\ref{eq:H_KG}) is an integral of motion, so its value $H_K$ (usually referred as the system's energy) remains constant and it also serves as a nonlinearity control parameter.

The second model is the DDNLS system, having the following Hamiltonian function
\begin{equation}
H_D = \sum_{l=1}^{N} \epsilon_l \lvert \psi_l \rvert ^2 +
\frac{\beta}{2}\lvert \psi_l \rvert^4 - \left(\psi_{l+1}\psi_l^{\ast} +
\psi_{l+1}^{\ast} \psi_l \right).
\label{eq:H_DDNLS}
\end{equation}
Here, $\psi_l$ is the complex wave function  at site $l$, $\beta \geq 0$ is the nonlinearity strength, $\epsilon_l$ are random parameters defining the on-site energy whose values are  chosen uniformly from
the interval $\left[- \frac{W}{2}, \frac{W}{2} \right]$, with $W$ denoting again the disorder strength. The canonical transformation $\psi_l = (q_l + ip_l)/\sqrt{2}$, $\psi_l^{\ast} = (q_l - ip_l)/\sqrt{2}$ brings \eqref{eq:H_DDNLS} to the form
\begin{equation}
H_D = \sum_{l=1}^{N} \frac{\epsilon _l}{2}(q_l ^2 + p_l ^2) + \frac{\beta}{8}(q_l ^2 +
p_l ^2)^2 - p_{l+1}p_l - q_{l+1}q_l,
\label{eq:H_DDNLS_qp}
\end{equation}
in which $q_l$ and $p_l$ are respectively the real valued generalized position and momentum at site
$l$. The corresponding Hamilton equations of motion
%$\dot{q}_l = \frac{\partial H_D}{\partial p_l}$,  $\dot{p}_l = - \frac{\partial H_D}{\partial q_l}$
take the form
\begin{equation}\label{eq:DDNLS_EM}
\begin{aligned}
      \dot{q}_l & =   p_l \left(\epsilon _l + \beta\frac{q_l ^ 2 + p_l ^2}{2} \right) - (p_{l-1} +
p_{l+1}), \\
      \dot{p}_l & =  -q_l  \left( \epsilon_l + \beta\frac{q_l ^ 2 + p_l ^2}{2} \right) + (q_{l -1} +
q_{l+1}).
\end{aligned}
\end{equation}
This set of equations conserves the total energy $H_D$ (\ref{eq:H_DDNLS_qp}) and the total norm of the system
\begin{equation}
\label{eq:norm}
    S = \sum_{l=1}^N \frac{1}{2} \left( q_l^2 + p_l^2 \right).
\end{equation}

In our study we follow the time evolution of initially localized excitations and analyze the characteristics of the induced wave packet propagations. We define  normalized energy distributions $\xi _l = \left[\frac{p_l ^2}{2} + \frac{\tilde{\epsilon}_l }{2} q_l^2 + \frac{q_l^4}{4} +
\frac{1}{4W} \left( q_{l + 1} - q_{l} \right)^2\right]/ H_K$ for the DKG model, while for the DDNLS system we consider normalized norm distributions $\xi _l=(q_l ^2 +
p_l ^2)/(2S)$. We compute the second moment $m_2 = \sum _l (l - \bar{l})^2 \xi_l$ of these distributions, which measures the distribution's extent  along with their participation number $P = 1/\sum _l \xi _l ^2$, which estimates the number of the strongest excited sites. In the definitions of these two quantities  $\bar{l} = \sum _l l \xi _l$ indicates the position of the distribution's center.

As a measure of the systems' chaoticity we  estimate the
maximum Lyapunov  exponent (mLE) $\Lambda_1$ as the limit for $t\rightarrow \infty$ of the finite-time mLE
\begin{equation}
\label{eq:ftMLE}
\Lambda (t) = \frac{1}{t}\ln \frac{\lvert \lvert\boldsymbol{w}(t)
\rvert \rvert}{\lvert \lvert\boldsymbol{w}(0) \rvert \rvert},
\end{equation}
i.e.~$\Lambda_1 = \lim _{t\to\infty} \Lambda(t)$. In (\ref{eq:ftMLE}) $\boldsymbol{w}(0)$ and $\boldsymbol{w}(t)$ are respectively phase space deviation vectors from the considered orbit at $t = 0$
and $t > 0$,  while $ \lvert \lvert \cdot \rvert \rvert$ denotes the usual Euclidian vector norm. The mLE is a widely used chaos indicator which measures the average rate of
growth (or shrinking) of a small perturbation to the solutions of dynamical
systems.
$\Lambda$ tends to zero for regular orbits following the power
law \cite{BGS76,S10}
\begin{equation}
\label{eq:LE_reg}
\Lambda \propto t^{-1},
\end{equation}
while it reaches some positive constant value for chaotic ones.

The time evolution of an initial deviation vector at time $t_0$ $\boldsymbol{w} (t_0) = \delta
 \boldsymbol{x} (t_0) = \left( \delta
 \boldsymbol{q} (t_0),  \delta
 \boldsymbol{p} (t_0) \right) =\left(\delta q_1 (t_0), \ldots, \delta q_N (t_0),
 \delta p_1 (t_0), \ldots, \delta p_N (t_0) \right)$ from a given  orbit with initial conditions $\boldsymbol{x} (t_0)=(\boldsymbol{q}(t_0),\boldsymbol{p}(t_0))$ is defined by the so-called variational equations (see for example \cite{S10} and references therein)
\begin{equation}
\label{eq:var_Hess}\boldsymbol{\dot{w}} (t) = \begin{bmatrix}
\dot{\delta q_l }(t) \\
\dot{\delta p_l}(t)
\end{bmatrix} =
\begin{bmatrix}
J_{2N} \boldsymbol{D}^2_{H} \left(\boldsymbol{x} (t) \right)
\end{bmatrix} \cdot \boldsymbol{w} (t_0), \,\,\, l=1,2,\ldots, N,
\end{equation}
where $J_{2N} = \begin{bmatrix}
0_N & I_N \\
-I_N & 0_N
\end{bmatrix}$, with $I_N$ and  $0_N$ being respectively  the identity and zero $N \times N$ matrices, while  $\boldsymbol{D}^2_{H} \left(\boldsymbol{x} (t) \right)$ is the
$2N\times 2N$ Hessian matrix with elements
$\left[\boldsymbol{D}^2_{H} \left(\boldsymbol{x} (t) \right)\right]_{i, j} =
\frac{\partial ^2 H}{\partial x_i \partial x_j } \bigg\rvert _{\boldsymbol{x}(t)}$, $i,j=1,2,\ldots, N$, evaluated at the reference orbit $\boldsymbol{x} (t)$. Equation (\ref{eq:var_Hess}) forms a set of linear equations with respect to $w_i(t)$, $i=1,2,\ldots, 2N$ (i.e.~the coordinates of vector $\boldsymbol{w}(t)$), whose  coefficients explicitly depend on
the time evolution of the reference orbit. Thus, the variational equations have to be  integrated simultaneously with the system's equations of motion.

We perform this task by implementing the so-called `tangent map method' \cite{SG10,GS11,GES12} using symplectic integration schemes. In particular, we integrate the DKG system by the two-part split ABA864 symplectic integrator of order four \cite{BCFLMM13}, and the DDNLS model by the sixth order symplectic scheme $ABC^{6}_{[SS]}$ \cite{SGBPE14,GMS16}, which is  based on the splitting of the DDNLS Hamiltonian in three integrable parts, as both integrators proved to be very efficient for these systems \cite{SGBPE14,GMS16,SS18}. Typically, we perform numerical simulations up to a final integration time of $t_f \approx 10^8$ time units. In order to exclude finite-size effects the number $N$ of lattice sites was increased up to $N \approx 7\,000$ in some of the considered cases. The used integration time steps $\tau \approx 0.18-0.5$ led to a very good conservation of the systems' integrals of motion, as the absolute energy relative error was usually kept smaller than $10^{-5}$ and the  absolute norm relative error of the DDNLS system was always below $10^{-3}$. For both models we imposed fixed boundary conditions $q_0=q_{N+1}=p_0=p_{N+1}=0$.

%==============================
\section{Numerical results}
\label{sec:num}

In our numerical simulations, we initially excite $L$ consecutive, central sites of the lattice. For the DKG model each of these $L$ sites gets the same amount of energy $\xi_l$
by setting $p_l = \pm \sqrt{2\xi_l}$ with randomly assigned signs, while all other sites have $p_l=0$. In addition, for all lattice sites we initially set $q_l=0$. In the DDNLS case each initially excited site gets a norm $\xi_l=1$ by putting $p_l = \pm \sqrt{2}$ with a random sign for each site. As in the case of the DKG model for all initially unexcited sites we set $p_l=0$, while we  put $q_l=0$ for all lattice sites. In the case of the DKG system the conserved quantity is the total energy $H_K=L\xi_l$, whose value does not depend on the choice of the considered disorder realization, i.e.~the fixed set of random values $\tilde{\epsilon}_l$, $l=1,2,\ldots,N$. As the DDNLS system conserves two quantities, the energy $H_D$ (\ref{eq:H_DDNLS_qp}) and the norm $S$ (\ref{eq:norm}), the above described choice of initial excitations sets the numerical value of the norm to $S=L$, while the exact value of $H_D$ depends on the implemented disorder realization $\epsilon_l$, $l=1,2,\ldots,N$, as well as the value of $\beta$.

In our analysis we consider several  weak and strong chaos cases and obtain statistical results of the behavior of a quantity $Q$ (e.g.~$m_2$, $P$, $\Lambda$) by averaging its values over 100 different disorder realizations and by smoothing these averaged values through a locally weighted difference algorithm \cite{CD88}. The outcome of this process will be denoted by  $\langle   Q \rangle$. Usually we present the time evolution of $Q$ in log-log scale and often estimate the related rate of change
\begin{equation}
\alpha_{Q} (\log_{10} t) = \frac{d\langle \log_{10}  Q \rangle}{d \log_{10} t},
\label{eq:aQ}
\end{equation}
through a central finite difference calculation, following the numerical process described in \cite{LBKSF10,BLSKF11}. We note that a practically constant value of $\alpha_Q$ indicates that the time evolution of $Q$ is described by the power law $Q \propto t^{\alpha_Q}$.

%==============================
\subsection{Lyapunov exponents}
\label{sec:Lyap}

We investigate the chaotic behavior of the DKG and the DDNLS systems by initially considering some parameter cases belonging to the weak chaos regime. In particular for the DKG system we study the following four cases:
\begin{description}
  \item[Case $W1_K$] $W=3$, $L=37$, $\xi_l = 0.01$;
  \item[Case $W2_K$] $W=4$, $L=1$, $\xi_l = 0.4$;
  \item[Case $W3_K$] $W=4$, $L=21$, $\xi_l = 0.01$.
  \item[Case $W4_K$] $W=5$, $L=13$, $\xi_l = 0.02$;
\end{description}
We also investigate the following four weak chaos cases of the DDNLS model:
\begin{description}
  \item[Case $W1_D$] $W=3$, $\beta = 0.03$, $L=21$, $\xi_l = 1$;
  \item[Case $W2_D$] $W=3$, $\beta = 0.6$, $L=1$, $\xi_l = 1$;
  \item[Case $W3_D$] $W=4$, $\beta = 1.0$, $L=1$, $\xi_l = 1$;
  \item[Case $W4_D$] $W=4$, $\beta = 0.04$, $L=21$, $\xi_l = 1$.
\end{description}

It is worth noting that DKG cases $W2_K$, $W3_K$ and $W1_K$ were also studied in \cite{SGF13} where they were named as cases I, II and III respectively. In that work averaged results over 50 disorder realizations for each case were presented, while here we increase the number of realizations to 100, improving in this way the statistical reliability of the obtained results. Let us also note that the parameter values of the DDNLS case $W4_D$ correspond to a well-known weak chaos case considered in \cite{LBKSF10,BLSKF11}.

\begin{figure}
\includegraphics[scale=0.68]{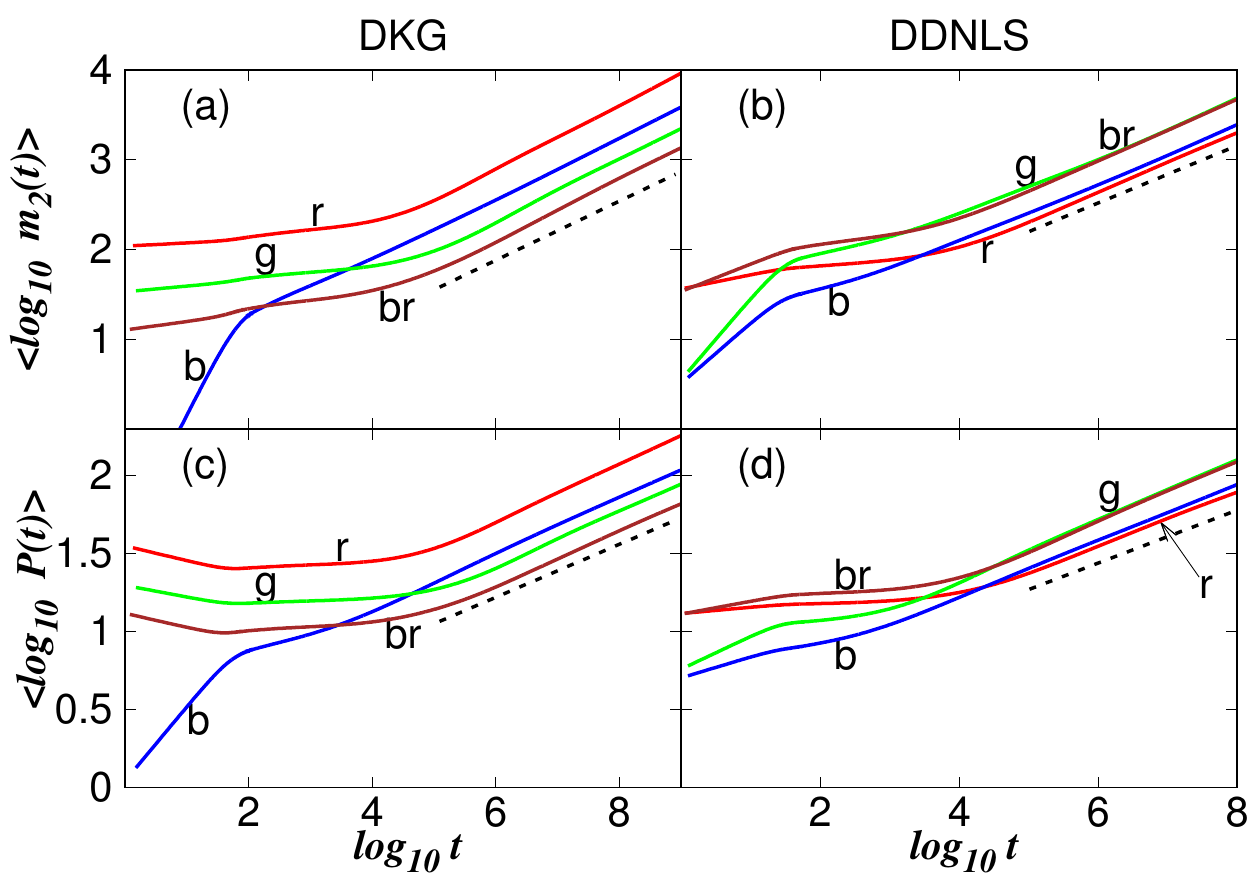}
  \caption{(Color online) Weak chaos. Averaged (and smoothed) results over 100 disorder realizations of the time evolution of the wave packets' second moment $m_2$ [(a), (b)] and  participation number $P$ [(c), (d)] for the DKG [(a), (c)] and the DDNLS [(b), (d)] systems. The straight dashed lines guide the eye for slopes $\frac{1}{3}$ [(a), (b)]  and $\frac{1}{6}$ [(c), (d)]. The presented cases are $W1_K$, $W2_K$, $W3_K$, $W4_K$  [(r) red; (b) blue; (g) green; (br) brown] for the DKG system and $W1_D$, $W2_D$, $W3_D$, $W4_D$ [(br) brown; (g) green; (b) blue; (r) red] for the DDNLS model. All panels are in log-log scale.}
\label{fig:weak_m2P}
\end{figure}

The results of Fig.~\ref{fig:weak_m2P} clearly verify that the considered DKG (left panels) and DDNLS cases (right panels) belong to the weak chaos spreading regime as the time evolution of $m_2$ (upper panels) and $P$ (lower panels) are well described by the power laws $m_2 \propto t^{1/3}$, $P \propto t^{1/6}$ in accordance to \cite{FKS09,SKKF09,LBKSF10,F10,BLSKF11}.
%%%%%%%%%%%%%%%%%%%%%%%%%%%%%%%%%%%%
%%%%%%%%%%%%%%%%%%%%%%%%%%%%%%%%%%%%

\begin{figure}
\includegraphics[scale=0.68]{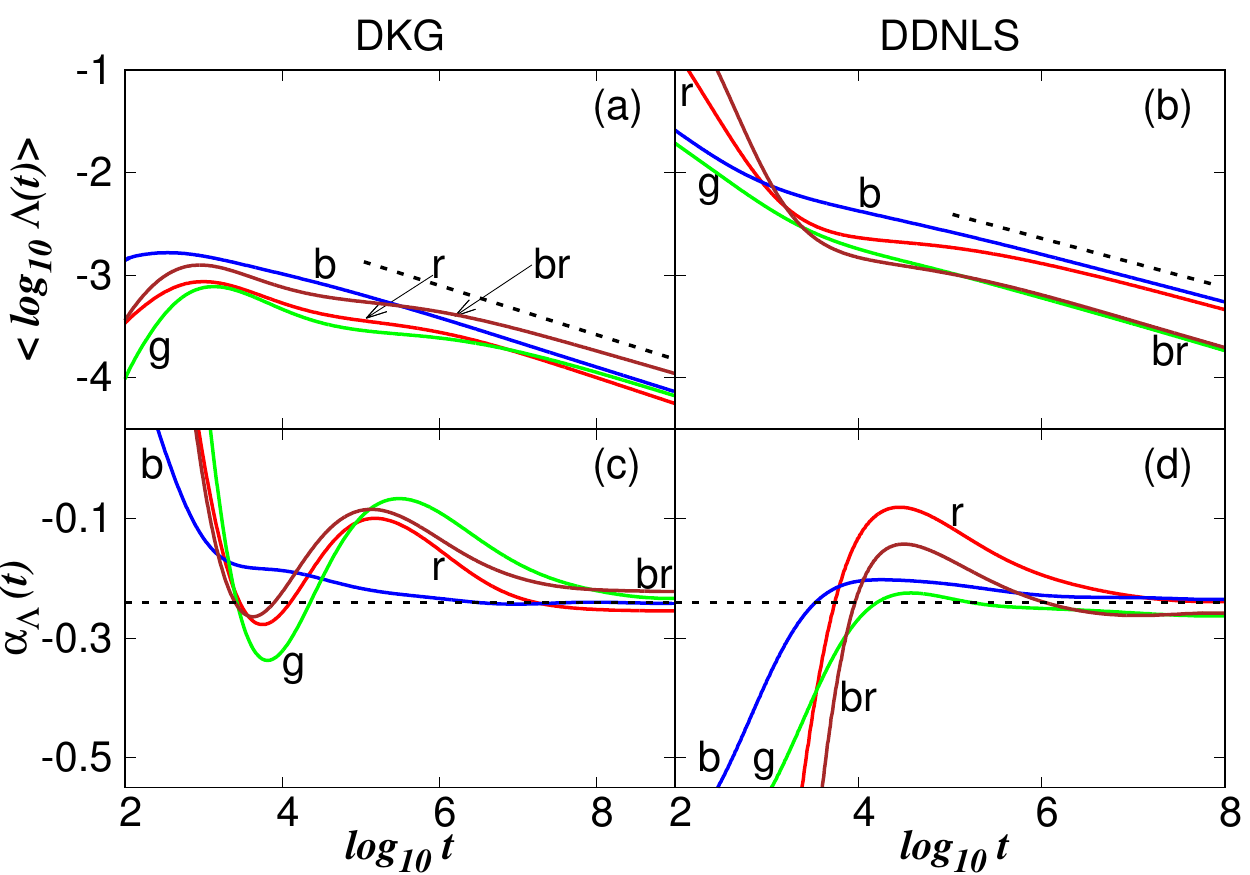}
  \caption{(Color online) Weak chaos. Averaged (and smoothed) results over 100 disorder realizations of the time evolution of the finite-time mLE $\Lambda (t)$ [(a), (b)] and the corresponding derivatives $\alpha_\Lambda$ (\ref{eq:aQ}) [(c), (d)]
  for the DKG [(a), (c)] and the DDNLS [(b), (d)] systems. The straight dashed lines indicate slopes $\alpha_\Lambda =-0.25$. The curve colors correspond to the cases presented in Fig.~\ref{fig:weak_m2P}. All panels are in log-log scale.}
\label{fig:weak_L}
\end{figure}

For all these weak chaos cases we compute in the upper panels of Fig.~\ref{fig:weak_L} the time evolution of the averaged over disorder realizations and smoothed $\Lambda$ for the DKG (left panel) and the DDNLS system (right panel). In the lower panels of  Fig.~\ref{fig:weak_L} we plot the numerically computed derivatives [see Eq.~(\ref{eq:aQ})] of the curves in the figure's upper panels. These results show that in all weak chaos cases the time evolution of the finite-time mLE converges toward the power law $\Lambda \propto t^{-0.25}$ \cite{com1}. This  is in agreement with the findings of \cite{SGF13} where the DKG cases $W2_K$, $W3_K$ and $W1_K$ were considered, while the extra case $W4_K$ studied here provides additional evidences of the validity of the $\Lambda$ power law decay. The new, important result here is that  this behavior is not restricted to the DKG model, but it is more general as it is also observed unaltered for the DDNLS model. This generality implies that the specific value of $\Lambda$'s decrease rate (i.e.~the exponent -0.25) characterizes the weak chaos regime.
%%%%%%%%%%%%%%%%%%%%%%%%%%%%%%%%%%%%
%%%%%%%%%%%%%%%%%%%%%%%%%%%%%%%%%%%%

As was extensively discussed in \cite{SGF13} the DKG system in the weak chaos regime becomes less chaotic in time since the value of $\Lambda$ follows a power law decay. This decrease of chaos strength can be understood in the following way. As the wave packet spreads the (constant) total energy is shared among more  activated degrees of freedom  as additional lattice sites are excited. Thus, the energy density of the  excited sites (which can be considered as the system's \emph{effective} nonlinearity strength) decreases. Nevertheless, the dynamics shows no signs of a crossover to regular behavior, which is characterized by $\Lambda \propto t^{-1}$, as the computed exponent $\alpha_\Lambda$ (lower panels of  Fig.~\ref{fig:weak_L}) saturates at  $\alpha_\Lambda \approx -0.25 \neq -1$. In a similar way to the DKG energy distribution, as the DDNLS norm distribution spreads the norm density of the excited sites decreases and consequently the nonlinear terms $\frac{\beta}{8}(q_l ^2 + p_l ^2)^2$ become weaker. Thus, the system becomes less chaotic and the value of $\Lambda$ decreases. Our results provide strong numerical evidences that this behavior is not a particularity of the DKG model, but it is quite general as it is manifested also in the DDNLS system, despite the fact that this system has two integrals of motion, the energy $H_D$ (\ref{eq:H_DDNLS_qp}) and the norm $S$ (\ref{eq:norm}).

Let us now turn our attention to the chaotic behavior of energy/norm propagations in the strong chaos spreading regime; an issue which was not considered  in \cite{SGF13}. As was explained in \cite{LBKSF10,F10,BLSKF11} the strong chaos subdiffusive regime can appear in cases of multi-site initial excitations. In this regime the dynamics is characterized by an initial faster, with respect to the weak chaos case, wave packet spreading, where $m_2 \propto t^{1/2}$ and $P \propto t^{1/4}$. This initial phase is followed by a subsequent slowing down of spreading, which asymptotically tends to the weak chaos behavior (i.e.~$m_2 \propto t^{1/3}$ and $P \propto t^{1/6}$).

In our study we consider six strong chaos parameter cases, three cases for the DKG model:
\begin{description}
  \item[Case $S1_K$] $W=2$, $L=83$, $\xi_l = 0.1$;
  \item[Case $S2_K$] $W=3$, $L=37$, $\xi_l = 0.1$;
  \item[Case $S3_K$] $W=3$, $L=83$, $\xi_l = 0.1$,
\end{description}
and three cases for the DDNLS system:
\begin{description}
  \item[Case $S1_D$] $W=3$, $\beta = 0.5$, $L=21$, $\xi_l = 1$;
  \item[Case $S2_D$] $W=3.5$, $\beta = 0.62$, $L=21$, $\xi_l = 1$;
  \item[Case $S3_D$] $W=3.5$, $\beta = 0.72$, $L=21$, $\xi_l = 1$.
\end{description}

\begin{figure}
\includegraphics[scale=0.68]{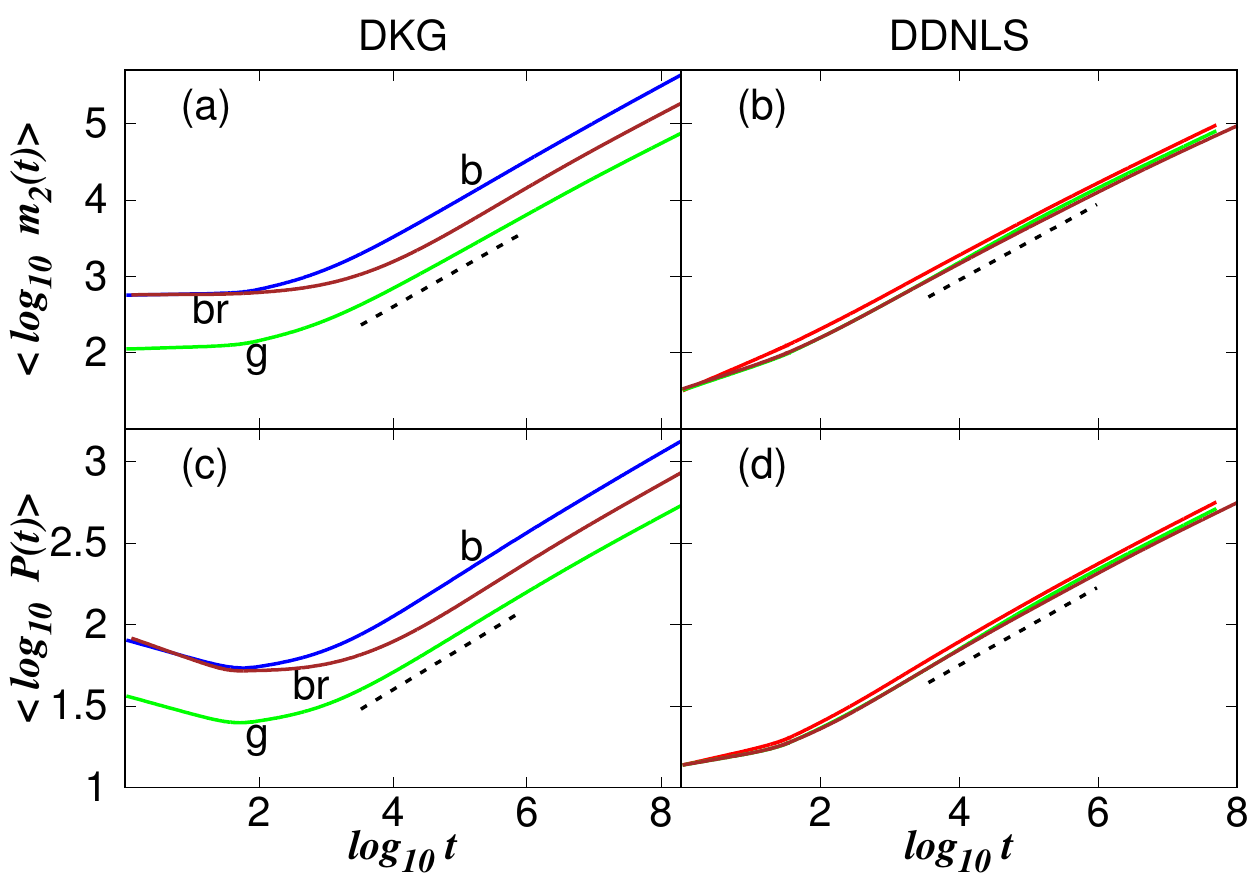}
  \caption{(Color online) Strong chaos. Similar to Fig.~\ref{fig:weak_m2P}. The straight dashed lines guide the eye for slopes $\frac{1}{2}$ [(a), (b)]  and $\frac{1}{4}$ [(c), (d)]. The presented cases are
  $S1_K$, $S2_K$,
  $S3_K$ [(b) blue; (g) green; (br) brown] for the DKG system [(a), (c)] and
  $S1_D$, $S2_D$, $S3_D$ [(g) green; (r) red; (br) brown] for the DDNLS model [(b), (d)].}
\label{fig:strong_m2P}
\end{figure}

\begin{figure}
\includegraphics[scale=0.68]{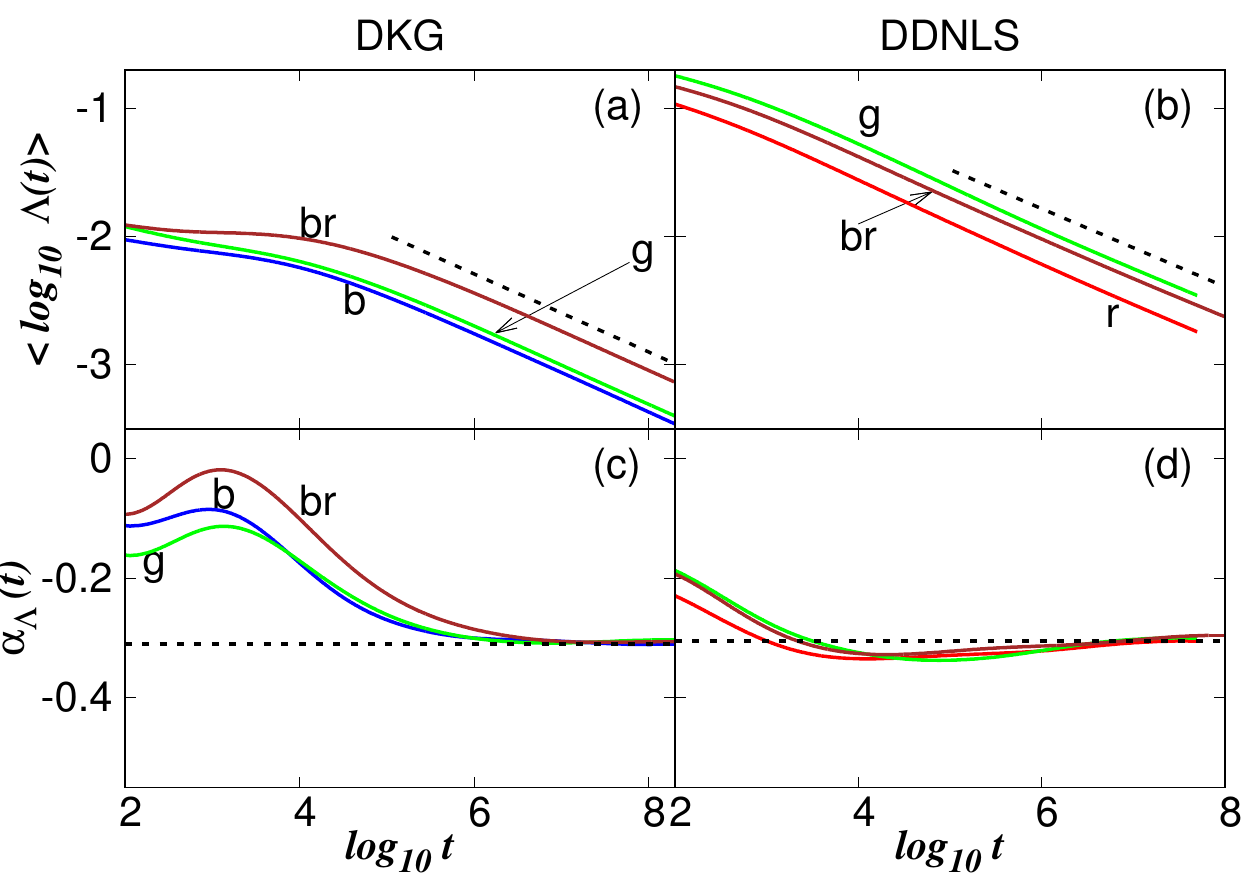}
  \caption{(Color online)  Strong chaos. Similar to Fig.~\ref{fig:weak_L}. The straight dashed lines indicate slopes $\alpha_\Lambda =-0.3$. The various curves correspond to the cases presented in Fig.~\ref{fig:strong_m2P}.}
\label{fig:strong_L}
\end{figure}

The results of Fig.~\ref{fig:strong_m2P} show that all these cases exhibit the characteristics of strong chaos, as $m_2 \propto t^{1/2}$ (upper panels) and $P \propto t^{1/4}$ (lower panels) for at least 2 decades,  for both the DKG (left panels) and the DDNLS model (right panels). This epoch is followed by a mild slowing down of the spreading process for $\log_{10}t \gtrsim 6$. The time evolution of $\Lambda$ in Fig.~\ref{fig:strong_L} shows a similar behavior to the one observed in the weak chaos case (Fig.~\ref{fig:weak_L}), i.e.~$\Lambda$ eventually decreases following a power law of the form  $\Lambda \propto t^{\alpha_{\Lambda}}$, without showing any signs of crossover to the law $\Lambda \propto t^{-1}$ and to regular dynamics. The difference is that now $\alpha_{\Lambda} \approx -0.3$ \cite{com2}, while in the weak chaos case we have $\alpha_{\Lambda} \approx -0.25$. The appearance of the value $\alpha_{\Lambda}=-0.3$ in  both models (lower panels of Fig.~\ref{fig:strong_L}) clearly shows the generality of this exponent, while its clear difference from the $\alpha_{\Lambda}=-0.25$ value observed in the weak chaos case is an additional indication of the dynamical differences of the two regimes.
%%%%%%%%%%%%%%%%%%%%%%%%%%%%%%%%%%%%
%%%%%%%%%%%%%%%%%%%%%%%%%%%%%%%%%%%%
%%%%%%%%%%%%%%%%%%%%%%%%%%%%%%%%%%%%
%%%%%%%%%%%%%%%%%%%%%%%%%%%%%%%%%%%%

As the strong chaos regime is a transient one, the evolution of $m_2$ and $P$ show signs of the crossover to the weak chaos dynamics, as their increase becomes slower  for $\log_{10} t \gtrsim 6$ (Fig.~\ref{fig:strong_m2P}). This happens because the values of $m_2$ and $P$ are determined by the current dynamical state of the wave packet. On the other hand, such changes are not visible in the evolution of $\Lambda$ (Fig.~\ref{fig:strong_L}). As the dynamics crosses over from the strong chaos behavior characterized by $\alpha_{\Lambda}=-0.3$ to the asymptotic weak chaos behavior associated with $\alpha_{\Lambda}=-0.25$, one would expect to see some  change in the values of $\alpha_{\Lambda}$ (lower panels of Fig.~\ref{fig:strong_L}) indicating this transition. Such changes are not observed because the value of $\Lambda$ (\ref{eq:ftMLE}) is influenced  by the whole evolution of the deviation vector [i.e.~the ratio $\lvert \lvert\boldsymbol{w}(t)
\rvert \rvert / \lvert \lvert\boldsymbol{w}(0) \rvert \rvert $ in (\ref{eq:ftMLE})] and consequently the whole history of the dynamics (which is predominately influenced by the strong chaos behavior),  and not from the current state of the systems. Thus, $\Lambda$ is not sensitive to subtle dynamical changes.  In the next section we will present some ways to capture such changes in the systems' chaotic behavior.

%==============================
\subsection{Deviation vector distributions}
\label{sec:DVD}

In order to analyze the dynamics of chaos evolution in the DKG and the DDNLS models we also compute the normalized DVD
\begin{equation}
\label{eq:DVD}
    \xi^D_l(t) = \frac{\delta q_l(t) ^2 + \delta p_l(t)^2}{\sum_l \left[ \delta q_l(t)^2 + \delta p_l(t)^2 \right]}, \,\,\, l=1,2,\ldots, N,
\end{equation}
created by the time evolution of the vector $\boldsymbol{w} (t)$ used for the computation of $\Lambda$  (\ref{eq:ftMLE}). Since $\boldsymbol{w} (t)$ eventually aligns to the most unstable direction in the system's phase space (which corresponds to the mLE), large $\xi^D_l$ values indicate at which lattice sites the sensitive dependence on initial conditions is higher. For this reason, such distributions were used in \cite{SGF13} to visualize the motion of chaotic seeds inside the spreading wave packet.

\begin{figure}
\includegraphics[scale=0.33]{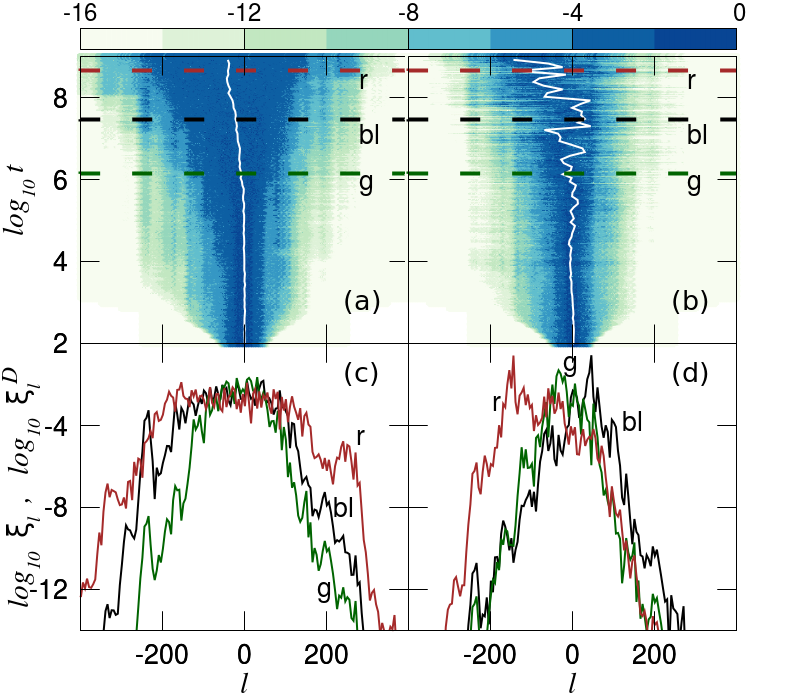}
  \caption{(Color online) DKG model, weak chaos. The dynamics of a representative initial condition of the $W1_K$ case for one disorder realization. Time evolution of (a) the normalized energy distribution $\xi_l$ and (b) the corresponding DVD. The color scales at the top of the figure are used for coloring lattice sites according to their $\log_{10} \xi_l$ (a) and $\log_{10} \xi_l^D$ (b) values. In both panels a white curve traces the distributions' center. Normalized energy distributions $\xi_l$ (c) and  DVDs  (d) at times $\log_{10}t=6.14$, $\log_{10}t=7.47$, $\log_{10}t=8.65$ [green (g); black (bl); red (r)]. These times are also denoted by similarly colored horizontal dashed lines in (a) and (b).}
\label{fig:palette_weak_kg}
\end{figure}

\begin{figure}
\includegraphics[scale=0.33]{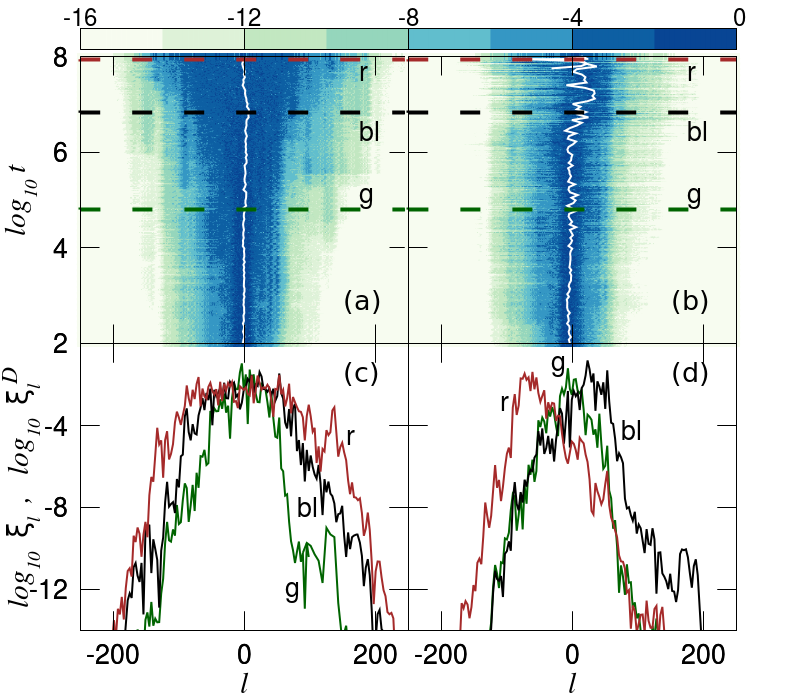}
  \caption{(Color online) DDNLS system, weak chaos. The dynamics of a representative initial excitation of the $W4_D$ case for one disorder realization. All panels are similar to the ones of Fig.~\ref{fig:palette_weak_kg}, with norm (instead of energy) distributions plotted in (a) and (c). The distribution snapshots in (c) and (d) are taken at times
  $\log_{10}t=4.8$, $\log_{10}t=6.82$, $\log_{10}t=7.94$ [green (g); black (bl); red (r)].}
\label{fig:palette_weak_ddnls}
\end{figure}

In Fig.~\ref{fig:palette_weak_kg}(a) [Fig.~\ref{fig:palette_weak_ddnls}(a)] we plot the time evolution of the energy density $\xi_l$ for the DKG system [norm density $\xi_l$ for the DDNLS model] for an individual set up belonging to the $W1_K$ [$W4_D$] weak chaos case, while in  Fig.~\ref{fig:palette_weak_kg}(b) [Fig.~\ref{fig:palette_weak_ddnls}(b)] the evolution of the corresponding DVD density is shown. In Figs.~\ref{fig:palette_weak_kg}(c), (d) [Fig.~\ref{fig:palette_weak_ddnls}(c), (d)] snapshots of these distributions taken at the instances denoted by horizontal dashed lines in  Figs.~\ref{fig:palette_weak_kg}(a),(b) [Fig.~\ref{fig:palette_weak_ddnls}(a), (b)] are shown.
%%%%%%%%%%%%%%%%%%%%%%%%%%%%%%%%%%%%
%%%%%%%%%%%%%%%%%%%%%%%%%%%%%%%%%%%%
%%%%%%%%%%%%%%%%%%%%%%%%%%%%%%%%%%%%
%%%%%%%%%%%%%%%%%%%%%%%%%%%%%%%%%%%%

From the results of Figs.~\ref{fig:palette_weak_kg} and \ref{fig:palette_weak_ddnls} we see that for both the DKG and the DDLNS models the energy/norm densities expand continuously to  larger regions of the lattice. This spreading is done more or less symmetrically around the position of the initial excitation as the evolution of the distributions' mean position [white curve in Figs.~\ref{fig:palette_weak_kg}(a) and \ref{fig:palette_weak_ddnls}(a)] is rather smooth, always  remaining close to the lattice's center. On the other hand, the DVDs, which stay always inside the excited part of the lattice, retain a more localized, pointy shape. At first the DVDs are located in the region of the initial excitation but they start moving around widely after $\log_{10}t \approx 6$, something which is clearly depicted in the time evolution of each DVD's mean position $\bar{l}_w = \sum _l l \xi_l^D$ [white curve in Figs.~\ref{fig:palette_weak_kg}(b) and \ref{fig:palette_weak_ddnls}(b)], as $\bar{l}_w$ shows random fluctuations with increasing amplitude. These results denote that the observed behavior (which was initially reported in \cite{SGF13} for the DKG system) is generic as it appears also for the DDNLS model. Based on such observations the authors of \cite{SGF13} used DVDs to represent the random motion of deterministic chaotic seeds inside the wave packet. These random oscillations of the chaotic seeds are essential in homogenizing chaos inside the wave packet, supporting in this way the wave packet's thermalization and subdiffusive spreading.

\begin{figure}
%\centerline{
%\hspace{1cm}
%\includegraphics[scale=1.10]{LE_fig_6_dvd_weak.eps}
%}
%\includegraphics[scale=0.8]{LE_fig_6_dvd_weak.eps}
%\includegraphics[scale=1.27]{LE_fig_7_new.eps}
\includegraphics[scale=1.27]{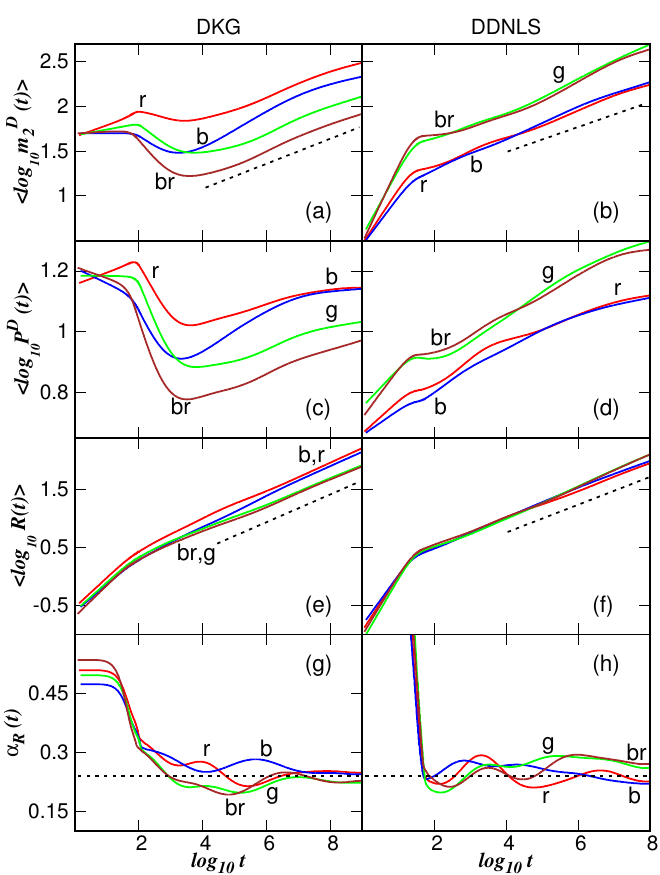}
  \caption{(Color online) DVD characteristics in the weak chaos regime. Time evolution of the averaged (and smoothed) over 100 disorder realizations, second moment $m_2^D$ [(a), (b)], participation number $P^D$ [(c), (d)] and $R$ (\ref{eq:R}) [(e), (f)]. The numerically computed derivatives $\alpha_R$ (\ref{eq:aQ}) of curves in (e) and (f) are respectively plotted in (g) and (h). Left panels contain results for the DKG model with curve colors corresponding to the cases presented in the left panels of Fig.~\ref{fig:weak_m2P}. Results for the DDNLS model are presented in the right panels with curve colors corresponding to the cases considered in the right panels of Fig.~\ref{fig:weak_m2P}. The straight dashed lines in (a) and (b) correspond to slope $0.14$, while in (e)-(h) indicate the slope $\alpha_R=0.24$. All horizontal axes are logarithmic. Panels (a)-(f) are in log-log scale.}
\label{fig:DVD_weak}
\end{figure}

For the created DVDs we also compute the time evolution of their  second moment $m_2^D$ and participation number $P^D$. Moreover, in order to quantify the range of the lattice region visited by the meandering localized DVD, we follow the evolution of the quantity
\begin{equation}
\label{eq:R}
    R(t)= \max_{[0,t]} \{ \bar{l}_w(t) \} - \min_{[0,t]} \{ \bar{l}_w(t) \}.
\end{equation}
The obtained results are presented in Fig.~\ref{fig:DVD_weak} for the weak chaos cases of both the DKG (left panels) and the DDNLS systems (right panels) considered in Sect.~\ref{sec:Lyap}. The DVDs' second moment [Figs.~\ref{fig:DVD_weak}(a), (b)] shows an asymptotic, slow growth ($m_2^D \propto t^{0.14}$), reaching values which are always smaller than the wave packets' $m_2$ [Figs.~\ref{fig:weak_m2P}(a), (b)] by at least one order of magnitude. The fact that the DVDs of Figs.~\ref{fig:palette_weak_kg} and \ref{fig:palette_weak_ddnls} retain a rather narrow, pointy shape remaining practically localized (although the place of their localization changes) is clearly reflected in their small and almost constant $P^D$ values [Figs.~\ref{fig:DVD_weak}(c), (d)]. For both the DKG and the DDNLS models $P^D$ attains small values (in the worst case of the order of $P^D \approx 20$ for $W2_D$) showing a tendency to asymptotically saturate to a constant number, since all curves of  Figs.~\ref{fig:DVD_weak}(c), (d) show signs of an eventual level off.
%%%%%%%%%%%%%%%%%%%%%%%%%%%%%%%%%%%%
%%%%%%%%%%%%%%%%%%%%%%%%%%%%%%%%%%%%

Thus, apart from the DVDs' profiles [Figs.~\ref{fig:palette_weak_kg}(b), (d) and \ref{fig:palette_weak_ddnls}(b), (d)], the slow increase of $m_2^D$ [Figs.~\ref{fig:DVD_weak}(a), (b)] and the practical constancy of $P^D$ [Figs.~\ref{fig:DVD_weak}(c), (d)] clearly show that the chaotic seeds retain a very localized character. Since the wave packet itself spreads continuously, the localized chaotic seeds, which constantly meander inside it, have to cover larger lattice regions as time increases. This becomes evident by the continuously  increasing values of $R$  (\ref{eq:R}) [Figs.~\ref{fig:DVD_weak}(e), (f)]. This increase is very well described, for both the DKG and the DDNLS models, by the power law $R \propto t^{\alpha_R}$ [Figs.~\ref{fig:DVD_weak}(e), (f)] with $\alpha_R \approx 0.24$ [Figs.~\ref{fig:DVD_weak}(g), (h)].

\begin{figure}
\includegraphics[scale=0.31]{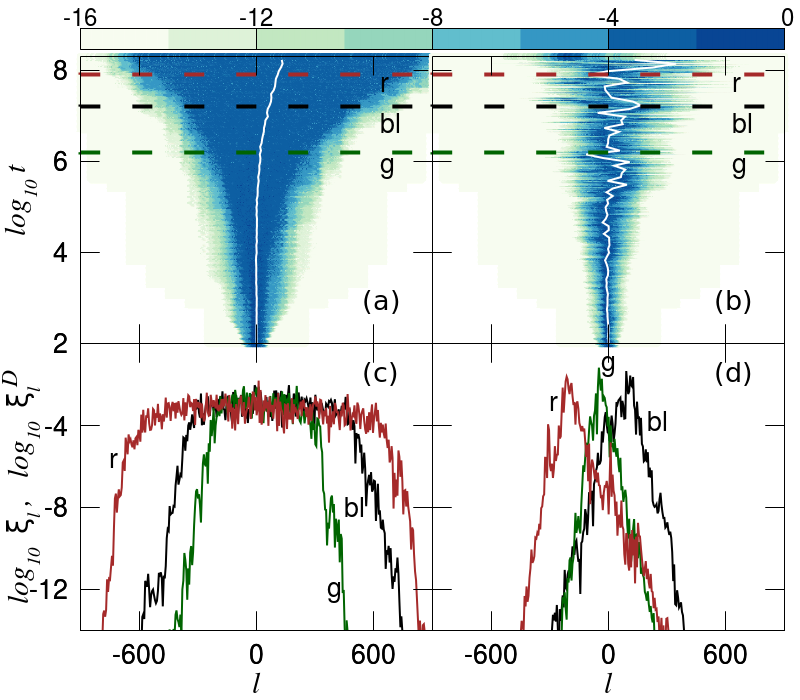}
  \caption{(Color online) DKG model, strong chaos. Similar to Fig.~\ref{fig:palette_weak_kg}, but for a representative initial condition of the $S3_K$ case. The distribution snapshots in the lower panels are taken at times $\log_{10}t=6.2$, $\log_{10}t=7.2$, $\log_{10}t=7.9$ [green (g); black (bl); red (r)].}
\label{fig:palette_strong_kg}
\end{figure}

\begin{figure}
\includegraphics[scale=0.31]{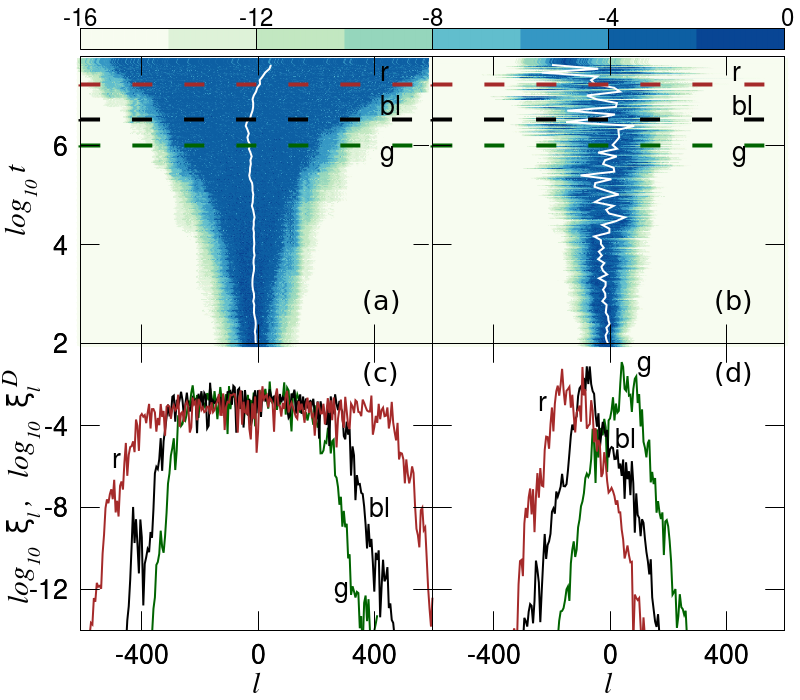}
  \caption{(Color online) DDNLS model, strong chaos. Similar to Fig.~\ref{fig:palette_weak_ddnls}, but for a representative initial condition of the $S3_D$ case. The distribution snapshots in the lower panels are taken at times $\log_{10}t=6.01$, $\log_{10}t=6.54$, $\log_{10}t=7.24$ [green (g); black (bl); red (r)].}
\label{fig:palette_strong_ddnls}
\end{figure}

Let us now investigate how chaotic seeds behave in the strong chaos regime. In Figs.~\ref{fig:palette_strong_kg}(a), (b) [Figs.~\ref{fig:palette_strong_ddnls}(a), (b)] we respectively plot the time evolution of the energy [norm] density and the corresponding DVD for an individual $S3_K$ [$S3_D$] set up, while snapshots of these distributions at some specific times are shown in Figs.~\ref{fig:palette_strong_kg}(c), (d) [Figs.~\ref{fig:palette_strong_ddnls}(c), (d)].
%%%%%%%%%%%%%%%%%%%%%%%%%%%%%%%%%%%%
%%%%%%%%%%%%%%%%%%%%%%%%%%%%%%%%%%%%
%%%%%%%%%%%%%%%%%%%%%%%%%%%%%%%%%%%%
%%%%%%%%%%%%%%%%%%%%%%%%%%%%%%%%%%%%

As in the weak chaos cases of Figs.~\ref{fig:palette_weak_kg} and \ref{fig:palette_weak_ddnls} the energy/norm density spreads smoothly and rather symmetrically around the lattice's center [Figs.~\ref{fig:palette_strong_kg}(a), (c) and \ref{fig:palette_strong_ddnls}(a), (c)], reaching sites further away with respect to the weak chaos cases [Figs.~\ref{fig:palette_weak_kg}(a), (c) and \ref{fig:palette_weak_ddnls}(a), (c)]. This is due to the fact that the strong chaos regime is characterized by a faster subdiffusive spreading than the one observed in the weak chaos case, which is reflected in the larger exponents in the power law increases of $m_2$ and $P$ (Figs.~\ref{fig:weak_m2P} and \ref{fig:strong_m2P}). On the other hand, the DVDs remain again localized, exhibiting fluctuations in their position, which appear earlier in time and have  larger amplitudes [Figs.~\ref{fig:palette_strong_kg}(b), (d) and Figs.~\ref{fig:palette_strong_ddnls}(b), (d)] with respect to the weak chaos case [Figs.~\ref{fig:palette_weak_kg}(b), (d) and Figs.~\ref{fig:palette_weak_ddnls}(b), (d)].

\begin{figure}
%\centerline{
%\hspace{1cm}
%\includegraphics[scale=1.10]{LE_fig_8_dvd_strong.eps}
%}
%\includegraphics[scale=0.8]{LE_fig_8_dvd_strong.eps}
%\includegraphics[scale=0.8]{LE_fig_10_new.eps}
\includegraphics[scale=1.27]{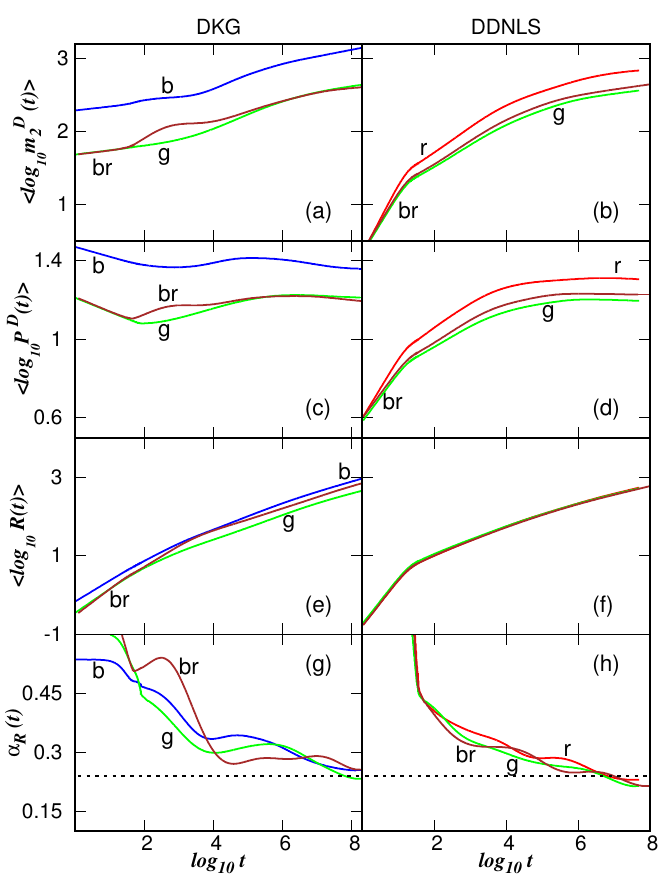}
 \caption{(Color online) DVD characteristics in the strong chaos regime. Similar to Fig.~\ref{fig:DVD_weak}, but for the strong chaos cases presented in Fig.~\ref{fig:strong_m2P}. The horizontal dashed lines in (g) and (h) indicate the slope $\alpha_R=0.24$ as in Figs.~\ref{fig:DVD_weak}(g), (h).}
\label{fig:DVD_strong}
\end{figure}

The DVDs' $m_2^D$ [Figs~\ref{fig:DVD_strong}(a), (b)] increases in time attaining larger values with respect to the weak chaos regime [Figs~\ref{fig:DVD_weak}(a), (b)], although this increase does not show signs of a constant rate (in log-log scale) as in the weak chaos case where $m_2^D \propto t^{0.14}$. In addition, a slowing down of the increase rate is observed at higher times especially for the DDNLS system [Figs~\ref{fig:DVD_strong}(b)]. The fact that the DVDs remain localized is depicted in the clear  tendency of their $P^D$ to saturate to values a little bit higher than the ones observed in the weak chaos case, as we get at most $P^D \approx 25$.
%%%%%%%%%%%%%%%%%%%%%%%%%%%%%%%%%%%%
%%%%%%%%%%%%%%%%%%%%%%%%%%%%%%%%%%%%

Since the wave packet spreads faster in the strong chaos case than in the weak chaos one, while the DVD remains again localized, one would expect  faster and wider movements of the chaotic seeds in order to achieve the wave packet's chaotization. The inspection of the $\bar{l}_w$ motion [white curves in Fig.~\ref{fig:palette_strong_kg}(b) and Fig.~\ref{fig:palette_strong_ddnls}(b)], as well as the evolution of $R$  (\ref{eq:R}) [Figs.~\ref{fig:DVD_strong}(e), (f)] and its derivative [Figs.~\ref{fig:DVD_strong}(g), (h)] show that this is true. $R$ grows faster than the $R \propto t^{0.24}$ increase observed in the weak chaos case [Figs.~\ref{fig:DVD_weak}(e), (f)], reaching also larger values by about one order of magnitude. The fact that the strong chaos regime is a transient one, as the dynamics will eventually crossover to the weak chaos spreading, is also reflected in the behavior of $R$ as its derivative $\alpha_R$ decreases in time [Figs.~\ref{fig:DVD_strong}(g), (h)], indicating the slowing down of the chaotic seeds' movement. For large times $\alpha_R$ show a tendency to reach values which are comparable to  the $\alpha_R=0.24$ [horizontal dashed line in Figs.~\ref{fig:DVD_strong}(g), (h)] seen in the weak chaos regime.

%==============================
\section{Summary and discussion}
\label{sec:sum}

We numerically investigated the chaotic behavior of one-dimensional nonlinear disordered lattices when Anderson localization is destroyed and spreading takes place.  In our study we considered two basic lattice models, which have been studied intensively in the last decade, the DKG and the DDNLS systems, and investigated their chaotic behavior in the weak and strong chaos spreading regimes. In particular, we performed extensive simulations of the chaotic propagation of initially localized excitations, for several weak and strong chaos parameter cases of these systems and obtained statistical results on ensembles of 100 disorder realizations in each case.

By computing the most commonly employed chaos indicator, the finite time mLE $\Lambda$, we provided clear evidences that although the chaoticity strength of the propagating wave packets decreases in time the dynamics retains its chaotic nature without any signs of a crossover to regular behavior. More specifically, we found that for both models and dynamical regimes $\Lambda$ decreases by following a power law  $\Lambda \propto t^{\alpha_{\Lambda}}$, which is characterized by $\alpha_{\Lambda}$ values different from  $\alpha_{\Lambda}=-1$  observed for regular motion. Moreover, the weak and strong chaos cases exhibit different $\alpha_{\Lambda}$ values, which remain the same for both studied systems, something which denotes the generality of these exponent values. In particular, we found that $\alpha_{\Lambda} \approx -0.25$ for the  weak chaos regime (in agreement to the results of \cite{SGF13}), while $\alpha_{\Lambda} \approx -0.3$ for the strong chaos regime. These particular values are related to the dynamical characteristics of each regime, but a theoretical explanation of this connection is still lacking.

Although the wave packet spreading remains chaotic, an important question is whether the wave packet's chaotization occurs fast enough to support its subdiffusive spreading. A way to tackle this question is by comparing the chaoticity time scale, which is usually called Lyapunov time $T_L$ (see for example \cite{S10} and references therein) and is estimated as
\begin{equation}
\label{eq:TL}
    T_L \sim \frac{1}{\Lambda},
\end{equation}
with some characteristic time scales related to the wave packet spreading. The latter can be done in two ways. Assuming that the spreading is characterized by an asymptotic momentary diffusion coefficient $D$, such that $m_2 \sim Dt$, then a characteristic spreading time scale $T_M$ can be obtained as
\begin{equation}
\label{eq:TM}
    T_M \sim \frac{1}{D}.
\end{equation}
Alternatively, one could define a  spreading time scale $T_P$ as the time required to increase the wave packet's participation number $P$ by one, so that
\begin{equation}
\label{eq:TP}
    T_P \sim \frac{1}{\dot{P}},
\end{equation}
with $\dot{P}$ being the time derivative of $P$.

For both the weak and the strong chaos regimes we have $m_2 \propto t^a$, $P \propto t^{a/2}$ \cite{FKS09,SKKF09,LBKSF10,F10,BLSKF11}, while our results show that $\Lambda \propto t^{\alpha_{\Lambda}}$. Then the ratios
\begin{equation}
\label{eq:T_ratios}
    \frac{T_M}{T_L}  \sim t^{1+\alpha_{\Lambda}-a}, \,\,\,\, \frac{T_P}{T_L}  \sim t^{1+\alpha_{\Lambda}-a/2}
\end{equation}
become
\begin{equation}
\label{eq:T_ratios_weak}
    \frac{T_M}{T_L}  \sim t^{\frac{5}{12}}, \,\,\,\, \frac{T_P}{T_L}  \sim t^{\frac{7}{12}},
\end{equation}
for the weak chaos regime, for which $a=1/3$ and $\alpha_{\Lambda}=-0.25$, and
\begin{equation}
\label{eq:T_ratios_strong}
    \frac{T_M}{T_L}  \sim t^{\frac{1}{5}}, \,\,\,\, \frac{T_P}{T_L}  \sim t^{\frac{9}{20}},
\end{equation}
for the strong chaos case characterized by $a=1/2$ and $\alpha_{\Lambda}=-0.3$. Thus, the chaoticity time scale $T_L$ remains always smaller than the spreading time scales $T_M$ and $T_P$, which implies that the wave packet's chaoticization is faster than its spreading.

The computation of the corresponding DVDs created by the deviation vector used to compute $\Lambda$ and of quantities related to their dynamics ($m_2^D$, $P^D$, $R$), allowed us to better capture the instantaneous features of the underlying chaotic behavior and to visualize the meandering motion of chaotic seeds inside the wave packet. In all studied cases the DVD retained a localized, pointy shape with its participation number $P^D$ remaining asymptotically constant to $P^D \approx 20-25$. As time increased the DVD exhibited oscillations of larger amplitudes in order to visit all regions inside the spreading wave packet. Consequently, the quantity $R$ (\ref{eq:R}), which tries to quantify the range of the lattice region visited by the DVD, increased in time. This increase is asymptotically characterized by a power law growth, $R\propto t^{0.24}$, in the weak chaos regime for both the DKG and the DDNLS systems. On the other hand, in the strong chaos case $R$ grows with a higher, but nonconstant, rate since the wave packet spreads faster than in the weak chaos case and the DVD visits a wider region. It is worth noting that this rate decreases in time, tending to the value $0.24$ observed in weak chaos regime. This is a direct consequence of the transient nature  of the strong chaos regime, as this regime eventually crosses over toward the weak chaos dynamics.

In conclusion, extending and completing previous results on the chaotic behavior of disordered lattices \cite{SGF13}, we numerically verified for both the DKG and the DDNLS model and the weak and strong chaos spreading regimes that (a) the deterministic chaoticity of wave packet dynamics persists in time, although its strength decreases,  (b) chaotic seeds meander inside the wave packet fast enough to ensure its chaotization, and (c) the characteristics of chaos evolution (like for example the power law $\Lambda \propto t^{\alpha_{\Lambda}}$) in the weak and strong chaos regimes are distinct for each case (e.g.~$\alpha_{\Lambda} \approx -0.25$ for weak chaos and $\alpha_{\Lambda} \approx -0.3$ for strong chaos), but also general as they are obtained for both  studied models.

An open question for future studies is the theoretical determination of the particular values of the exponent $\alpha_{\Lambda}$ for each dynamical regime. Another interesting problem is the investigation of the chaotic behavior of disordered lattices of higher dimensionality, in the spirit of the studies presented here. Some first, preliminary investigations (see Fig.~4(e) of \cite{SS18}) showed that in the weak chaos regime of a two-dimensional DKG system $\Lambda$  decreases to zero by following a power law which is again different than the $t^{-1}$ law observed for regular motion. We expect to perform in the  near future a more systematic study of such questions for both the weak and strong chaos regimes in various models of two-dimensional disordered lattices.

%==============================
\begin{acknowledgments}
We thank S.~Flach for useful discussions. Ch.~S.~and B.~M.~M.~were supported by the National Research Foundation of South Africa (Incentive Funding for Rated Researchers, IFFR and Competitive Programme for Rated Researchers, CPRR). B.~S.~was partially funded by the University of Cape Town International and Refugee Grant, as well as the Muni University AfDB-HEST staff development fund. The authors would like to thank the  High Performance Computing facility of the University of Cape Town (\url{http://hpc.uct.ac.za}) and the Center for High Performance Computing (\url{https://www.chpc.ac.za}) for the provided computational resources needed for performing the largest part of this paper's  computations, as well as their user-support teams for their help on many practical issues. We also thank the two anonymous referees for their comments, which helped us improve the presentation of our work.
\end{acknowledgments}

%==============================


\begin{thebibliography}{99}

\bibitem{A58} P.~W.~Anderson, Phys.~Rev.~{\bf 109}, 1492
  (1958).

\bibitem{WBLR97} D.~S.~Wiersma, P.~Bartolini, A.~Lagendijk, and R.~Righini,
Nature {\bf 390}, 671 (1997).

\bibitem{CSG00} A.~A.~Chabanov, M.~Stoytchev, and A.~Z.~Genack,
Nature {\bf 404}, 850 (2000).

\bibitem{RZ03} E.~Runge and R.~Zimmermann, Lect.~Notes Phys.~{\bf 630}, 145 (2003).

\bibitem{GC05} A.~Z.~Genack and A.~A.~Chabanov, J.~Phys.~A  {\bf 38}, 10465 (2005).

\bibitem{SGAM06} M.~St\"{o}rzer, P.~Gross, C.~M.~Aegerter, and G.~Maret,
Phys.~Rev.~Lett.~{\bf 96}, 063904 (2006).

\bibitem{BJZBHLCSBA08} J.~Billy, V.~Josse, Z.~Zuo, A.~Bernard, B.~Hambrecht,
  P.~Lugan, D.~Cl\'ement, L.~Sanchez-Palencia, P.~Bouyer, and A.~Aspect,
  Nature {\bf 453}, 891 (2008).

\bibitem{HSPSV08}
H.~Hu, A.~Strybulevych, J.~H.~Page, S.~E.~Skipetrov,  and B.~A.~van Tiggelen,
Nature Phys.~{\bf 4}, 945 (2008).

\bibitem{KMZD11} S.~S.~Kondov, W.~R.~McGehee, J.~J.~Zirbel,  and B.~DeMarco,
Science {\bf 334}, 66 (2011).

\bibitem{KKFA08} G.~Kopidakis, S.~Komineas, S.~Flach and S.~Aubry, Phys.~Rev.~Lett.~{\bf 100}, 084103 (2008).

\bibitem{PS08} A.~S.~Pikovsky  and D.~L.~Shepelyansky, Phys.~Rev.~Lett.~{\bf 100},  094101 (2008).

\bibitem{FKS09} S.~Flach, D.~O.~Krimer, and Ch.~Skokos, Phys.~Rev.~Lett.~{\bf 102}, 024101 (2009).

\bibitem{SKKF09}
 Ch.~Skokos, D.~O.~Krimer, S.~Komineas and S.~Flach, Phys.~Rev.~E {\bf 79}, 056211 (2009).

\bibitem{GS09}  I.~Garc\'{i}a-Mata and D.~L.~Shepelyansky, Phys.~Rev.~E {\bf 79},
026205 (2009).

\bibitem{VKF09} H.~Veksler, Y.~Krivolapov,  and S.~Fishman, Phys.~Rev.~E {\bf 80}, 037201 (2009).

\bibitem{MAPS09} M.~Mulansky, K.~Ahnert, A.~Pikovsky, and D.~L.~Shepelyansky, Phys.~Rev.~E, \textbf{80}, 056212 (2009).

%\bibitem{AS09} S.~Aubry and R.~Schilling, Physica D {\bf 238},
%  2045 (2009).

\bibitem{MP10}    M.~Mulansky and A.~Pikovsky, Europhys.~Lett.~{\bf 90}, 10015 (2010).

\bibitem{LBKSF10} T.~V.~Laptyeva, J.~D.~Bodyfelt,  D.~O.~Krimer, Ch.~Skokos, and S.~Flach, Europhys.~Lett.~{\bf 91}, 30001 (2010).

\bibitem{SF10} Ch.~Skokos and S.~Flach, Phys.~Rev.~E {\bf 82}, 016208 (2010).

\bibitem{KF10} D.~O.~Krimer and S.~Flach, Phys.~Rev.~E {\bf 82},
  046221 (2010).

\bibitem{F10} S.~Flach, Chem.~Phys.~{\bf 375}, 548 (2010).

\bibitem{JKA10} M.~Johansson, G.~Kopidakis, and S.~Aubry, Europhys.~Lett.~{\bf 91}, 50001 (2010).

\bibitem{B11} D.~M.~Basko, Ann.~Phys.~(N.Y.) {\bf 326}, 1577 (2011).

\bibitem{BLSKF11} J.~D.~Bodyfelt, T.~V.~Laptyeva, Ch.~Skokos, D.~O.~Krimer, and S.~Flach, Phys.~Rev.~E {\bf 84}, 016205 (2011).

\bibitem{BLGKSF11}    J.~D.~Bodyfelt, T.~V.~Laptyeva., G.~Gligoric, D.~O.~Krimer, Ch.~Skokos, and S.~Flach, Int.~J.~Bifurcation Chaos {\bf 21}, 2107 (2011).

\bibitem{A11} S.~Aubry, Int.~J.~Bifurcation Chaos, {\bf 21}, 2125
  (2011).

\bibitem{ILF11} M.~V.~Ivanchenko, T.~V.~Laptyeva, and
  S.~Flach, Phys.~Rev.~Lett.~{\bf 107}, 240602 (2011).

\bibitem{MLT12} M.~I.~Molina, N.~Lazarides, and G.~P.~Tsironis, Phys.~Rev.~E {\bf 85}, 017601 (2012).

\bibitem{VG12} B.~Vermersch and J.~C.~Garreau, Phys.~Rev.~E {\bf 85}, 046213 (2012).

\bibitem{MF12} E.~Michaely and S.~Fishman, Phys.~Rev.~E {\bf
    85}, 046218 (2012).

%\bibitem{MF12b} E.~Michaely and S.~Fishman, Eur.~Phys.~J.~B {\bf
%    85}, 362 (2012).

\bibitem{B12} D.~M.~Basko, Phys.~Rev.~E {\bf 86}, 036202 (2012).

\bibitem{MP12}    M.~Mulansky and A.~Pikovsky, Phys.~Rev.~E {\bf 86}, 056214 (2012).

\bibitem{LBF12} T.~V.~Laptyeva, J.~D.~Bodyfelt, and S.~Flach, Europhys.~Lett.~{\bf 98}, 60002 (2012).

\bibitem{MI12} A.~V.~Milovanov, and A.~Iomin, Europhys.~Lett.~{\bf 100}, 10006 (2012).

%\bibitem{RRF13} K.~Rayanov, G.~Radons, and S.~Flach,
%  Phys.~Rev.~E {\bf 88}, 012901 (2013).

%\bibitem{VG13} B.~Vermersch and J.~C.~Garreau, Eur.~Phys.~J.~Spec.~Top.~{\bf 217} 109 (2013).

\bibitem{LTDMIM13}
E.~Lucioni, L.~Tanzi, C.~D'Errico, M.~Moratti, M.~Inguscio, and G. Modugno,   Phys.~Rev.~E {\bf 87}, 042922  (2013).

\bibitem{VG13b} B.~Vermersch and J.~C.~Garreau, New
  J.~Phys.~{\bf 15} 045030 (2013).

\bibitem{MP13} M.~Mulansky and A.~Pikovsky, New J.~Phys.~{\bf 15}, 053015 (2013).

\bibitem{SGF13} Ch.~Skokos, I.~Gkolias, and S.~Flach, Phys.~Rev.~Lett.~{\bf 111}, 064101 (2013).

%\bibitem{GRF13} G.~Gligoric, K.~Rayanov, and S.~Flach, Europhys.~Lett.~{\bf
%    101}, 10011 (2013).

\bibitem{LBF13} T.~V.~Laptyeva, J.~D.~Bodyfelt, and S.~Flach,
  Physica D {\bf 256}, 1 (2013).

\bibitem{TSL14}
O.~Tieleman, Ch.~Skokos, and A.~Lazarides A., Europhys.~Lett.~{\bf 105}, 20001 (2014).

\bibitem{ILF14} M.~V.~Ivanchenko, T.~V.~Laptyeva, and S.~Flach, Phys.~Rev.~B {\bf 89}, 060301(R) (2014).

\bibitem{ABSD14}    Ch.~Antonopoulos, T.~Bountis, Ch.~Skokos, and L.~Drossos,  Chaos {\bf 24}, 024405 (2014).

\bibitem{ES14} L.~Ermann and D.~L.~Shepelyansky, J.~Phys.~A {\bf 47} 335101 (2014).

\bibitem{LIF14} T.~V.~Laptyeva, M.V.~Ivanchenko, and S.~Flach, J.~Phys.~A {\bf 47} 493001 (2014).

\bibitem{B14} D.~M.~Basko, Phys.~Rev.~E {\bf 89}, 022921 (2014).

\bibitem{F15} S.~Flach, Lect.~Notes Math.~{\bf 2146}, 1 (2015).

\bibitem{MKP16} A.~J.~Mart\'{i}nez, P.~G.~Kevrekidis, and M.~A.~Porter, Phys.~Rev.~E {\bf 93}, 022902 (2016).

\bibitem{ATS16} V.~Achilleos, G.~Theocharis, and Ch.~Skokos,
Phys.~Rev.~E {\bf 93}, 022903 (2016).

\bibitem{ASB17}    Ch.~Antonopoulos, Ch.~Skokos, and T.~Bountis,   Chaos Solitons Fractals {\bf 104}, 129 (2017).

\bibitem{I17}  A.~Iomin, Comput.~Math.~Applic.~{\bf 73}, 914 (2017).

\bibitem{ATS18} V.~Achilleos, G.~Theocharis, and Ch.~Skokos,
Phys.~Rev.~E {\bf 97}, 042220 (2018).

\bibitem{SBFS07} T.~Schwartz, G.~Bartal, S.~Fishman, and M.~Segev,
  Nature {\bf 446}, 52 (2007).

\bibitem{RDFFFZMMI08} G.~Roati, C.~D'Errico, L.~Fallani, M.~Fattori,
  C.~Fort, M.~Zaccanti, G.~Modugno, M.~Modugno, and M.~Inguscio,
  Nature {\bf 453}, 895 (2008).

\bibitem{LAPSMCS} Y.~Lahini, A.~Avidan, F.~Pozzi, M.~Sorel, R.~Morandotti,
  D.~N.~Christodoulides, and Y.~Silberberg, Phys.~Rev.~Lett.~{\bf 100}, 013906 (2008).

\bibitem{LDTRZMLDIM11}
E.~Lucioni, B.~Deissler, L.~Tanzi, G.~Roati, M.~Zaccanti, M.~Modugno, M.~Larcher, F.~Dalfovo, M.~Inguscio, and G. Modugno,   Phys.~Rev.~Lett.~{\bf 106}, 230403  (2011).

\bibitem{M98} M.~I.~Molina,  Phys.~Rev.~B {\bf  58}, 12547 (1998).

\bibitem{BGGS80a} G.~Benettin, L.~Galgani, A.~Giorgilli, and
  J.-M.~Strelcyn, Meccanica {\bf
    15}, 9 (1980).

\bibitem{BGGS80b} G.~Benettin, L.~Galgani, A.~Giorgilli, and
  J.-M.~Strelcyn, Meccanica {\bf
    15}, 21 (1980).

\bibitem{S10} Ch.~Skokos, Lect.~Notes Phys.~{\bf
    790}, 63 (2010).

\bibitem{BGS76}
  G.~Benettin, L.~Galgani, and J.-M.~Strelcyn,  Phys.~Rev.~A {\bf 14},  2338  (1976).

\bibitem{SG10} Ch.~Skokos and E.~Gerlach,  Phys.~Rev.~E, {\bf 82}, 036704 (2010).

\bibitem{GS11} E.~Gerlach and Ch.~Skokos, Discr.~Cont.~Dyn.~Sys.-Supp. 2011, 475

\bibitem{GES12} E.~Gerlach, S.~Eggl and Ch.~Skokos, Int.~J.~Bifurcation Chaos {\bf 22}, 1250216 (2012).

\bibitem{BCFLMM13} S.~Blanes, F.~Casas, A.~Farres, J.~Laskar,
  J.~Makazaga, and  A.~Murua, App.~Num.~Math.~{\bf 68}, 58 (2013).

\bibitem{SGBPE14} Ch.~Skokos, E.~Gerlach, J.~D.~Bodyfelt,
  G.~Papamikos, and S.~Eggl, Phys.~Lett.~A {\bf 378}, 1809 (2014).

\bibitem{GMS16} E.~Gerlach, J.~Meichsner, and Ch.~Skokos,
  Eur.~Phys.~J.~Spec.~Top.~{\bf 225}, 1103 (2016).

\bibitem{SS18}   B.~Senyange, Ch.~Skokos, Eur.~Phys.~J.~Spec.~Top., ~{\bf 227}, 625 (2018).

\bibitem{CD88} W.~S.~Cleveland and S.~J.~Devlin, \textit{J. Am. Stat.
    Assoc.} \textbf{83}, 596 (1988).

\bibitem{com1} In particular, averaging the values of $\alpha_{\Lambda}$ obtained from the smoothed values of  $\langle \log_{10}  \Lambda \rangle$, in the last decade of Figs.~\ref{fig:weak_L}(c) and (d) we get $-0.2539 \pm 0.0003$ ($W1_K$), $-0.2412 \pm 0.0005 $ ($W2_K$), $-0.231 \pm 0.003$ ($W3_K$), $-0.2216 \pm 0.0004$ ($W4_K$), $-0.259 \pm 0.001$ ($W1_D$), $-0.2621 \pm 0.0009$ ($W2_D$), $-0.2346 \pm 0.0008$ ($W3_D$) and $-0.238 \pm 0.001$ ($W4_D$).

\bibitem{com2} In a similar way as in \cite{com1}, from the results of  Figs.~\ref{fig:strong_L}(c) and (d) we get  $\alpha_{\Lambda} = -0.3104 \pm 0.0005$ ($S1_K$), $-0.3038 \pm 0.0008$ ($S2_K$), $-0.3063 \pm 0.003$ ($S3_K$), $-0.3002 \pm 0.0008$ ($S1_D$), $-0.3056 \pm 0.0009$ ($S2_D$) and $-0.297 \pm 0.001$ ($S2_D$).

%\bibitem{XXXX} XXXXXXXXXXXXXX

%\bibitem{schulte2006cold}
%  T. Schulte, S. Drenkelforth, J. Kruse, W. Ertmer, J. Arlt, A. Kantian, L Sanchez-Palencia, L. Santos, An. Sanpera, K. Sacha, and others,
%  Acta Phys. Pol. A
%  1,
%  109,
%  89--99
%  (2006)

% \bibitem{cleveland1988locally}
%  S. W. Cleveland and S. J. Devlin,
%  J Am Stat Assoc.
%  83,
%  403,
%  596--610
%  (1988)



\end{thebibliography}
\end{document}